\input{aipcheck}

\documentclass[
    ,final            
  ]
  {aipproc}

\layoutstyle{6x9}

\begin{document}

\title{Monitoring of Bright Blazars with MAGIC in the 2007/2008 Season}

\classification{95.55.Ka, 95.85.Pw, 98.54.Cm}
\keywords      {Active Galactic Nuclei: individual; BL Lacertae objects:
individual; gamma-rays: observations; gamma-ray telescopes}
\author{Konstancja Satalecka$^{\star,1}$
    Ching-Cheng Hsu,$^{\star,2}$
    Elisa Bernardini,$^1$ \\
    Giacomo Bonnoli,$^3$
    Nicola Galante,$^2$
    Florian Goebel,$^{2,\dagger}$
    Elina Lindfors,$^4$ \\
    Pratik Majumdar,$^1$
    Antonio Stamerra$^3$ and
    Robert Wagner$^2$ \\
    on behalf of the MAGIC collaboration}
{
address={$^1$ DESY, Platanenallee 6, D-15738 Zeuthen, Germany, \\
    $^2$ Max-Planck-Institut f\"ur Physik, F\"ohringer Ring 6, D-80805
M\"unchen, Germany, \\
    $^3$ Dipartimento di Fisica, Universit\`a degli Studi di Siena,
Via Roma 56, I-53100 Siena, Italy and \\
    $^4$ Tuorla Observatory, Dept.\ Physics and Astronomy, University
of Turku, FI-20014 Turku, Finland. \\
    $^\star$ Corresponding authors. Email: konstancja.satalecka@desy.de, cchsu@mppmu.mpg.de \\
    $^\dagger$ deceased}
}

\begin{abstract}
Because of the short duty-cycles and observation-time constraints, studies of
bright TeV (E>100\,GeV) blazars are mostly restricted to flaring episodes or
rather short (days to few weeks) multiwavelength campaigns. At the same time,
long-term studies of these objects are essential to gain a more complete
understanding of the blazar phenomenon and to constrain theoretical models
concerning jet physics. Only unbiased long-term studies are adequate for the
determination of flaring state probabilities and for estimating the statistical
significance of possible correlations between TeV flaring states and other
wavebands or observables, such as neutrino events. Regular observations also
provide triggers for multiwavelength ToO observations originating from the TeV
waveband. These are particularly needed to identify and study orphan
TeV flares, i.e. flares without counterparts in other wavebands. In 2007/8 the
MAGIC telescope has monitored three TeV blazars on a regular basis: Mrk 501,
Mrk 421, and 1ES 1959+650. We present preliminary results of these observations
including the measured light curves and a correlation study for VHE 
$\gamma$-rays and X-rays and VHE $\gamma$-rays and optical R-band for Mrk 421.
\end{abstract}

\maketitle


\section{Introduction}
With the new generation of Imaging Atmospheric Cerenkov Telescopes (IACTs) 
like HESS \cite{Aharonian:2006pe}, MAGIC \cite{Albert:2007xh} 
and VERITAS \cite{Acciari:2008ah}  
Very High Energy (VHE) $\gamma$-ray astronomy 
became a very dynamic and exciting discipline with many new detections of 
galactic and extra-galactic objects every year.
Almost one third of the objects detected in VHE $\gamma$-rays
are blazars, i.e. Active Galactic Nuclei (AGNs) which contain relativistic jets
pointing approximately in the direction of the observer. Their energy spectra 
show no or very weak emission lines, but a continuous distribution with two 
broad peaks: one in the UV to soft X-ray band and a second one in the GeV-TeV 
range. One of the most interesting aspects of blazars is their variability, 
observed in all frequencies and on different time scales ranging from weeks to 
minutes \cite{Aharonian:2007ig} \cite{Albert:2007zd}.

The data collected so far in many multiwavelenght observations
is not yet enough to fully constrain the theoretical 
models which try to explain the acceleration and emission mechanisms in 
blazars. In particular it is not yet clear if the leptonic or hadronic 
processes play a decisive role. For example, the Synchrotron-Self Compton 
(SSC) leptonic models \cite{Tavecchio:1998xw} can successfully describe most 
of the existing data and offer a reasonable explanation of the fast 
variability of blazars. 
On the other hand, the hadronic models, like the 
Synchrotron Mirror Model (SMM) \cite{Reimer:2005sj} or Synchrotron Proton 
Blazar (SPB) \cite{Muecke:2002bi} model can also very well describe the SED 
structure. Additionally they explain the ''orphan'' gamma-ray flares (like 
the one observed for 1ES 1959+650 in 2002 \cite{Krawczynski:2003fq}) and 
predict emission of high energy neutrinos. 

\section{AGN monitoring}
A long term monitoring of the VHE $\gamma$-ray flux variability performed 
in the GeV-TeV energy range will give vital input to understand the problems 
mentioned above. 

Observations scheduled independently of any knowledge of the
source state provide us with an unbiased distribution of the flux states.
Many of the previously performed measurements were triggered by an observed
enhancement of flux state in other wavebands and therefore
observations of AGNs during a low flux state are still sparse. Consequently
any statistical study which requires high statistics on various flux levels
is difficult. An example of such a study might be the determination of flaring 
state probabilities, essential for estimation of 
the statistical significance of possible correlations between flaring states 
and other observables, such as neutrino events \cite{Satalecka:2007}. Here a long term 
monitoring plays a crucial role, especially in a view of the results expected 
from the IceCube neutrino observatory \cite{Ahrens:2003ix}.

Another interesting application is the investigation of spectral changes during
periods of different source activity, which can improve our knowledge about the
acceleration and emission processes. 

Last but not least AGN monitoring can serve to trigger Target of Opportunity 
(ToO) observations. The ToO observations may be performed by the IACT issuing 
the ToO trigger but may also include other IACTs - allowing to increase 
the time coverage of the observations - or telescopes
and satellites observing other wavelengths. X-ray obsrvations are especially 
interesting in a context of the ''orphan'' TeV flares.

\section{The MAGIC Telescope}

MAGIC is currently the largest single-dish IACT 
for VHE $\gamma$-ray astronomy. It is located on 
the Canary Island of La Palma, at an altitude of 2200\,m a.s.l and has been 
in scientific operation since summer 2004. 
The camera is equipped with 576 photo-multiplier 
tubes (PMTs). In January 2007 a major upgrade of the MAGIC Telescope with 
a Multiplexed Fiber-Optic 2\,GSamples/s FADC Data Acquisition system took place
\cite{Goebel:2007cb}. The fast readout minimizes the influence of the background
from the light of the night sky and the additional information on the time 
structure of the event helps to reduce the background from hadronic 
showers \cite{Aliu:2008pd}. The trigger threshold of MAGIC is around 60\,GeV\footnote{The trigger threshold is defined as the peak of the energy distribution of the triggered events.}. 
A source emitting $\gamma$-rays 
at a flux level of 1.6\% of the Crab Nebula can be detected with 5\,$\sigma$ 
significance within 50 hours of observation time. The sensitivity is sufficient
to establish a flux level of about 30\% of the Crab flux above 300\,GeV for a 
30\,min observation (Fig.1). A quick on-line analysis system allows to estimate
the flux level of the observed source and facilitate the issuing of ToO triggers 
during data taking. 
MAGIC can also observe under moderate moon and twilight conditions with only 
slightly lower sensitivity.
The construction of a second telescope (MAGIC-II) is being finalized 
and the new telescope is undergoing various tests. It is planned
to start stereoscopic observations in spring 2009.
\begin{figure}
  \includegraphics[height=.4\textheight]{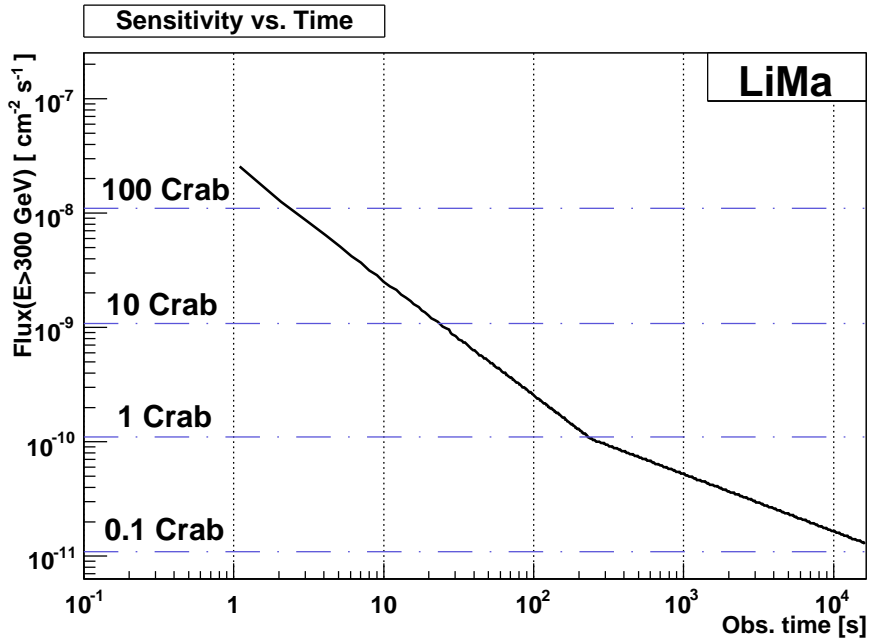}
  \caption{MAGIC sensitivity \cite{LiMa:1983} as a function of exposure time.}
\end{figure}

\section{Monitoring strategy}
Due to the low duty cycle (10-12\%) the operation time of IACTs is very 
precious and the observations have to be scheduled very carefully in order to
achieve the right balance between the high priority, deep observations and AGN 
monitoring.
Therefore usually previous generation IACTs which are still operational are  
used for this purpose. For example the Whipple\footnote{http://veritas.sao.arizona.edu/content/blogsection/6/40/} 
telescope, which is leading a monitoring program since 2005 for five well 
established TeV sources, or DWARF, one of the former HEGRA telescopes 
currently being refurbished and brought back into operation on La Palma \cite{DWARF}. In the case of MAGIC, a high sensitivity latest generation
IACT, some of the monitoring observations can be performed 
during moderate moonlight or twilight, which keeps the impact on the overall
observation schedule low. 

In order to achieve a dense sampling, up to 40 short observations per source
are scheduled, evenly distributed over the observable time during MAGIC 
observation period. Each of them should be long enough to detect a given minimum 
flux level taking into account the sensitivity of the telescope. In the case 
of MAGIC, three sources were chosen for a regular monitoring: Mrk 421, Mrk 501 
and 1ES 1959+650. The first two are relatively bright and usually 15-30\,min 
observations are scheduled for them. 1ES~1959+650 being fainter requires 
longer observation times, at least 30 minutes per single exposure.

\section{Results}
In this section we will present the preliminary results of the MAGIC AGN 
monitoring program for the observation season 2007/2008. The data have been 
processed with the standard MAGIC analysis tools \cite{Albert:2007xh}. 
A fraction of the data has been removed due to poor observation conditions.
All cuts are optimized and verified with Crab data 
in the corresponding observational seasons.
\subsection{Mrk 421}
Mrk 421 is the first extragalactic object detected at energies above 500\,GeV 
\cite{Punch:1992} \cite{Petry:1996yj} and one of the best studied.  
Here the preliminary results of the data which were 
taken from February 2007 to June 2008 are presented (82 hours). 
After the data selection, about 66 hours of data 
(80\%) were used for further analysis. The observations were mostly 
performed in wobble mode, which allows to simultaneously collect signal 
and background events.

\begin{figure}
  \includegraphics[width=1.\textwidth]{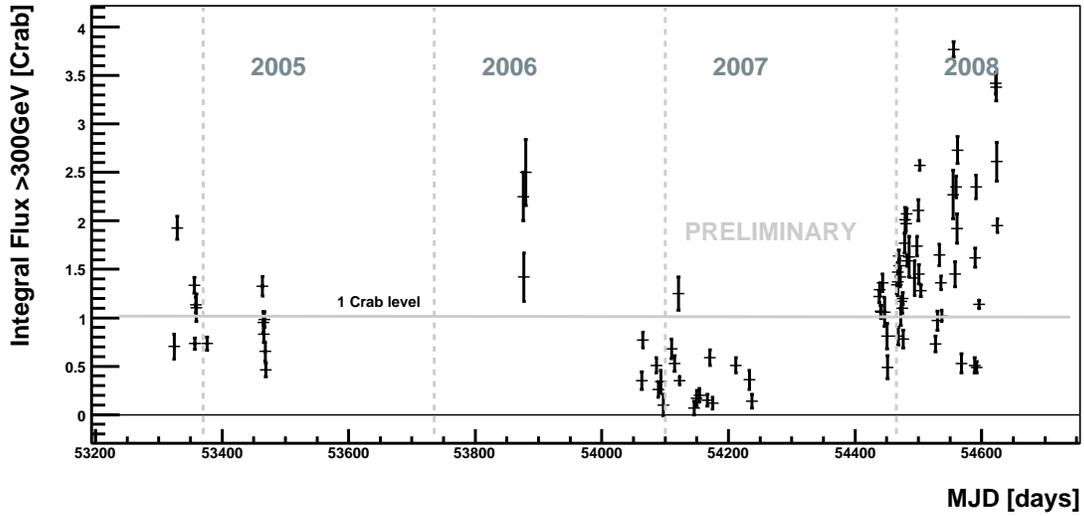}
  \caption{\label{LC_Mrk421_all}
Mrk 421 light curve showing all data collected by MAGIC \cite{Albert:2006jd} \cite{Goebel:2007uu}.
}
\end{figure}

\begin{figure}
  \includegraphics[width=1.\textwidth]{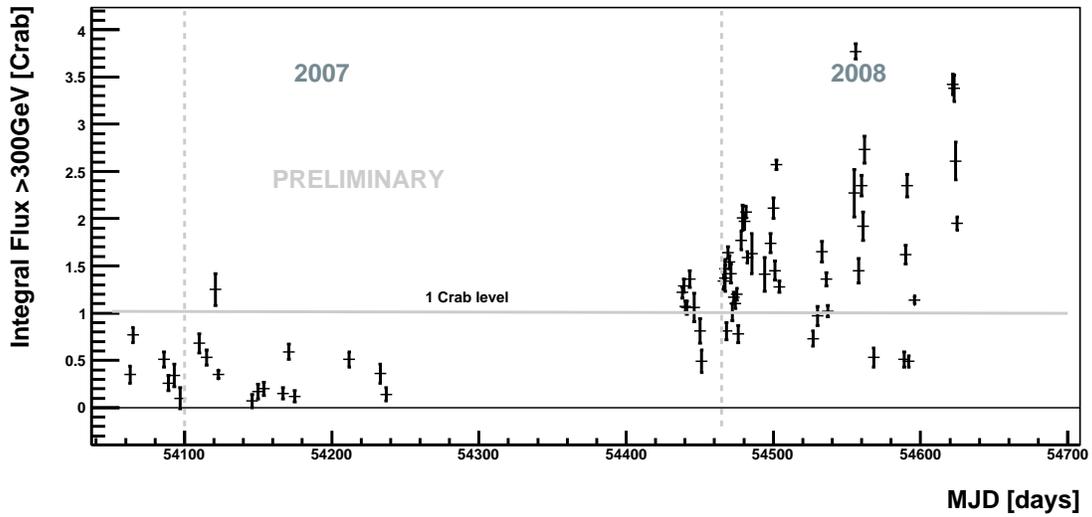}
  \caption{\label{LC_Mrk421_0708}
Mrk 421 light curve showing MAGIC data from AGN monitoring season 2007/2008.
}
\end{figure}

Fig.\ref{LC_Mrk421_all} shows all the data collected from Mrk 421 by 
the MAGIC telescope since 2004 \cite{Albert:2006jd} \cite{Goebel:2007uu}.
In Fig.\ref{LC_Mrk421_0708} the 2007/2008 observational period is shown magnified.
We can clearly see that Mrk 421 was especially active in 2008 and
the flux was never below 0.5\,Crab. During this time of high activity, 
in accordance 
with the ToO agreement, MAGIC issued several alerts to HESS and VERITAS. 
Detailed analysis of the collected data, such as intra-night variability and 
spectrum measurement will be published elsewhere.

In the SSC framework, a correlation between TeV photon and the other 
wavelengths, especially X-ray, is expected \cite{Maraschi:1999pg}. 
Since usually the TeV 
data is relatively poorly sampled, it is not easy to get really simultaneous 
data, we therefore searched for X-ray and optical measurements which are 
within 6\,h before or after the TeV observations.
 
\begin{figure}
  \includegraphics[height=.5\textheight]{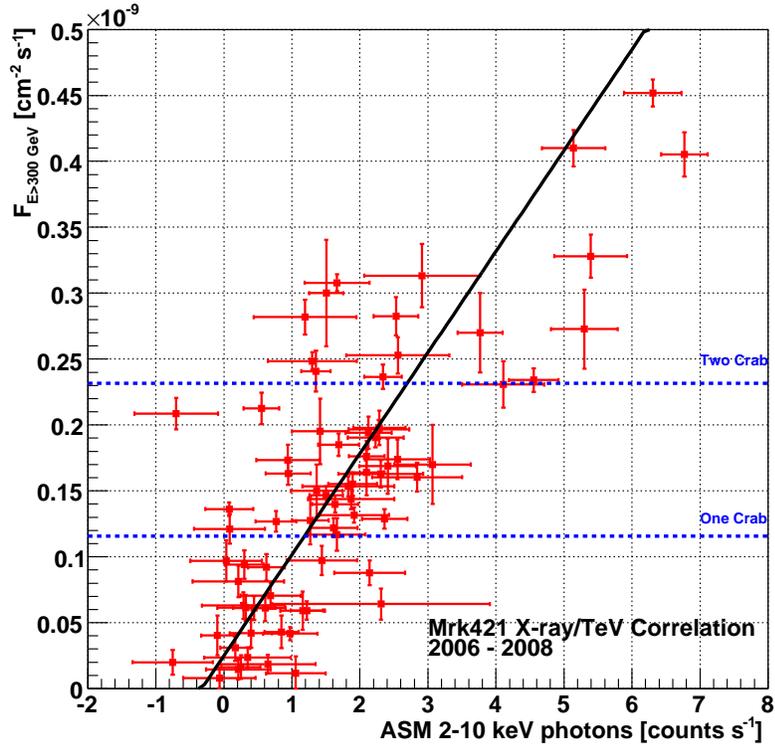}
  \caption{\label{TeVXray}VHE $\gamma$-ray (MAGIC) - X-ray (ASM) correlation plot for Mrk 421.}
\end{figure}

In this work we use the X-ray data taken by the RXTE/ASM satellite, which is 
designed for all sky monitoring observations. The data are available 
on the web page\footnote{http://xte.mit.edu/ASM\_lc.html} immediately after the 
observations. The averages and errors of ASM data points are calculated 
on a dwell-by-dwell\footnote{one dwell lasts 90 seconds} data basis. 
If the number of dwells are fewer than five, 
we discard that data point; 75 TeV-X-ray measurements pairs were 
finally selected (shown in Fig.\ref{TeVXray}). 

\begin{figure}
  \includegraphics[height=.5\textheight]{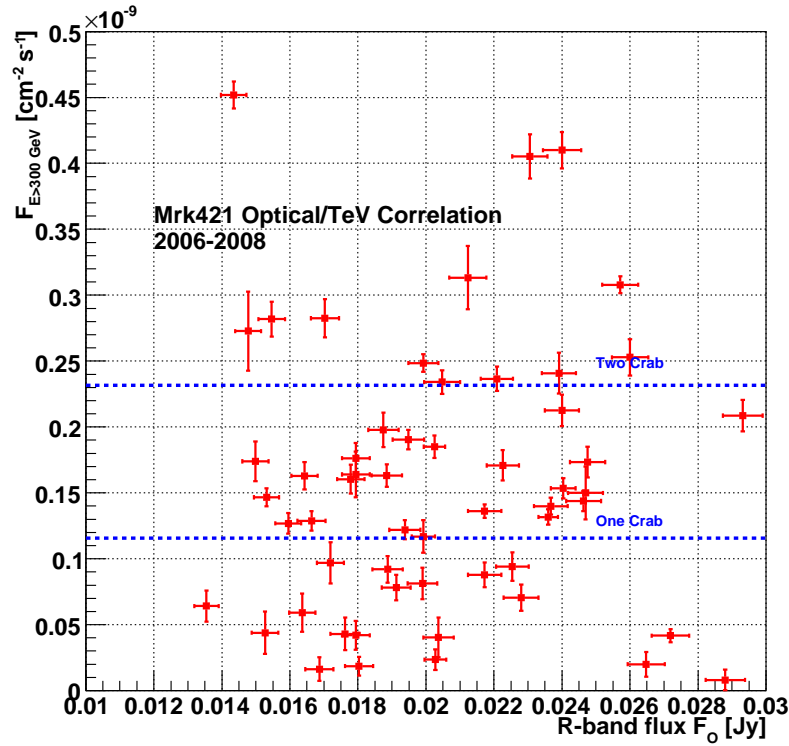}
  \caption{\label{TeVoptical}VHE $\gamma$-ray (MAGIC) - R-band (Tuorla) correlation plot for Mrk 421.}
\end{figure}

The optical R-band data were provided by the Tuorla Observatory 
Blazar Monitoring Program\footnote{http://users.utu.fi/kani/1m/} with the 
1.03\,m telescope at the Tuorla Observatory Finland and the 35\,cm KVA telescope 
at La Palma, Canary Islands. From the optical flux the flux of the host galaxy 
and the flux contribution from the companion galaxies were substracted 
\cite{Nilsson:2007ax}. Eventually, 56 measurements pairs were found 
(Fig.\ref{TeVoptical}). 

The correlation coefficients for both plots were calculated using Pearson's method
(see section 14.5 in \cite{638765}). 
For X-ray/TeV correlation, the correlation coefficient value is r~ = 0.77$\pm$0.05. 
For optical/TeV plot, the coefficient value is r = 0.03$\pm$0.14.
We thus found a significant correlation with X-rays. 
This type of correlation is easier accommodated in leptonic models 
(see e.g. \cite{Wagner:2008cw} and references therein) but there also exist 
certain hadronic models which support such correlation (see e.g. 
\cite{Muecke:2002bi} or \cite{Aharonian:2000pv}). 
However, one has to keep in mind, that the data were not taken strictly 
simultaneously and any conclusion as to the origin or mechanisms of the emission 
should be made very carefuly.

\subsection{Mrk 501}
The BL Lac object Mrk 501 was the second established TeV-blazar 
\cite{Quinn:1996dj} \cite{Bradbury:1996mi}. MAGIC has been observing this source 
since the year 2005 \cite{Albert:2007zd}. Here we present preliminary results 
based on an observation time of 16 hours (after quality selection), collected 
between February 2007 and August 2008. In order to maximize the time coverage 
of this source, observations were carried out mostly in the presence of moderate 
moonlight or twilight (9 hours, i.e. 56\% of the observation time). 
The observations were performed in wobble mode. 

\begin{figure}
  \includegraphics[width=1.\textwidth]{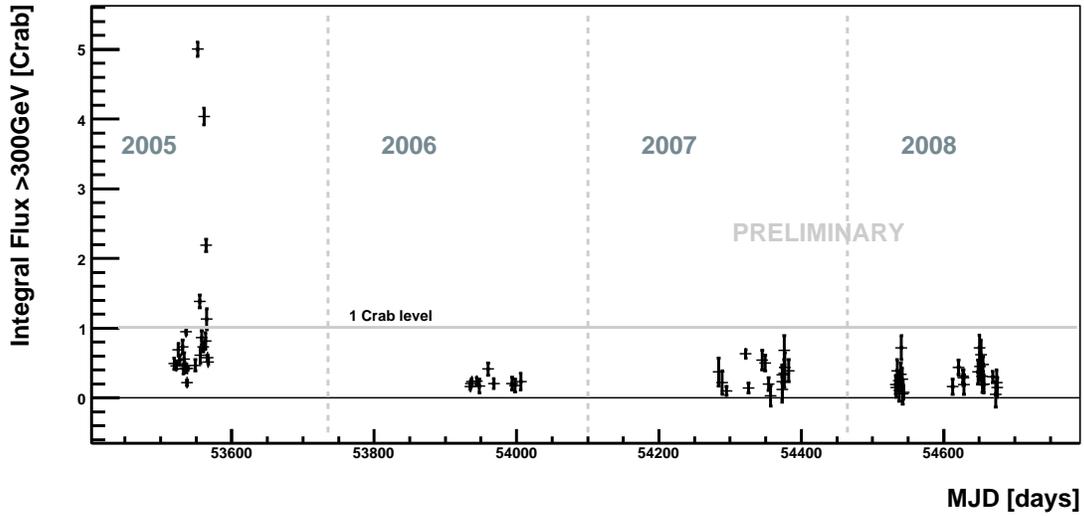}
  \caption{\label{LC_Mrk501_all}
Mrk 501 light curve showing all data collected by MAGIC \cite{Goebel:2007uu} \cite{Albert:2007zd}.
}
\end{figure}

\begin{figure}
  \includegraphics[width=1.\textwidth]{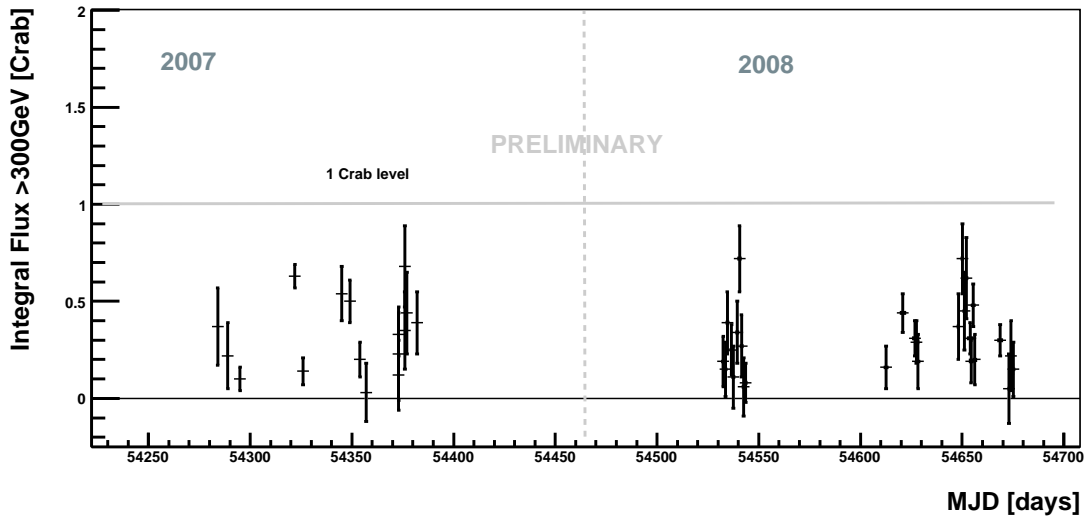}
  \caption{\label{LC_Mrk501_0708}
Mrk 501 light curve showing MAGIC data from AGN monitoring season 2007/2008.
}
\end{figure}

Fig. \ref{LC_Mrk501_all} shows all the data collected form Mrk 501 by the MAGIC
telescope since 2005  \cite{Albert:2007zd} \cite{Goebel:2007uu}. 
In Fig. \ref{LC_Mrk501_0708} the 2007/2008 monitoring season is shown magnified. 
During this period the source was in a relatively low state 
(below 1\,Crab). On the other hand a dense sampling was obtained and a 
statistical analysis, e.g. estimation of the of the source state probability, 
based on the above data is in progress.

\subsection{1ES 1959+650}
The observation of an ''orphan'' TeV flare (i.e. not accompanied by a 
corresponding X-ray
flare) in 2002 \cite{Krawczynski:2003fq} made 1ES 1959+650 one of
the most interesting VHE $\gamma$-ray sources. 
MAGIC has been observing this object
since 2004 \cite{Aliu:2005jv}, a regular monitoring was first performed 
in the 2005/2006 season \cite{Goebel:2007uu}. 
Unfortunately, the source position allows observations only under a relatively 
high zenith angles (35-50\,deg). 
Moreover, when not in a high state the source is rather faint compared to 
those discussed above.

MAGIC observed 1ES 1959+650 from April 2007 till October 2008 for 27 hours 
(including 13 hours of data taken during moonlight or twilight). 
Observations were performed in wobble mode. After rejection of data due to poor
quality 3.4 hours of data taken in 2008 were analyzed. The overall significance 
of this sample is 2.5\,$\sigma$, which allows us to set an upper limit on the flux 
above 300\,GeV of 2.53\,$\times 10^{-11}$\,ph/cm$^2$/s at 90\% C.L. Comparing to 
previous results \cite{Aharonian:2003} \cite{Gutierrez:2006ak} we can state 
that MAGIC observed this source in its usual low state.  
\section{Conclusions}
During the observational season 2007/2008 MAGIC monitored three sources: 
Mrk 421, Mrk 501 and 1ES 1959+650. The detailed analysis of the collected data 
is going on. In this work we presented preliminary results of the measured 
flux levels for all three sources. We also investigated the possible 
correlation of the TeV and X-ray/optical flux levels for Mrk 421. We found a 
significant correlation with X-rays and no correlation with optical R-band. 
The most interesting material for further studies are the 
many fast flares observed for Mrk 421 in 2008. Mrk 501 and 1ES 1959+650 were 
found in low state, but dense sampling of Mrk 501 flux states looks promising 
for statistical studies.
  

\begin{theacknowledgments}
  We would like to thank the Instituto de Astrofisica de Canarias for the 
excellent working conditions at the Observatorio del Roque de los Muchachos 
in La Palma. The support of the German BMBF and MPG, the Italian INFN and 
Spanish MCINN is gratefully acknowledged. This work was also supported by 
ETH Research Grant TH 34/043, by the Polish MNiSzW Grant N N203 390834, 
and by the YIP of the Helmholtz Gemeinschaft.
\end{theacknowledgments}



\bibliographystyle{aipproc}   

\bibliography{AGN_monitoring}

\IfFileExists{\jobname.bbl}{}
 {\typeout{}
  \typeout{******************************************}
  \typeout{** Please run "bibtex \jobname" to optain}
  \typeout{** the bibliography and then re-run LaTeX}
  \typeout{** twice to fix the references!}
  \typeout{******************************************}
  \typeout{}
 }

\end{document}